\documentclass[aps,showpacs,preprintnumbers,amsmath, amssymb]{revtex4}

\usepackage{float}
\usepackage{graphics,epsfig}
\usepackage{graphicx}
\usepackage{dcolumn}
\usepackage{bm}

\begin{document}
\baselineskip=0.8 cm

\title{{\bf Upper bound on the radii of regular ultra-compact star photonspheres }}
\author{Yan Peng$^{1}$\footnote{yanpengphy@163.com}}
\affiliation{\\$^{1}$ School of Mathematical Sciences, Qufu Normal University, Qufu, Shandong 273165, China}

\vspace*{0.2cm}
\begin{abstract}
\baselineskip=0.6 cm
\begin{center}
{\bf Abstract}
\end{center}

We investigate the photonsphere in the background of
regular asymptotically flat compact stars.
The analysis includes the general hairy compact star considering
the matter fields' backreaction on the metric
in various gravity theories.
We prove that the photonsphere of the compact star has
an upper bound expressed in terms of the ADM mass of the spacetime.
In the case of negative isotropic trace,
a stronger upper bound can be obtained.

\end{abstract}

\maketitle
\newpage
\vspace*{0.2cm}

\section{Introduction}

General relativity predicts that null bound geodesics may exist outside
compact objects, such as black holes and horizonless compact stars \cite{c1,c2}.
The null geodesics can provide valuable information about the structure and geometry of the curved spacetime.
In particular, the circular null geodesics on which photon can orbit the central compact object
are known as photonspheres. The photonspheres are of great importance from the astrophysical
and theoretical aspects \cite{ad1,ad2,ad3,ad4,ad5,c3,c4,c5,z1,z2,z3}.

On the side astrophysics, it was found that photonspheres play an important role in
the optical appearance of a compact star
to external observers in the asymptotic region.
For example, the strong gravitational lensing phenomenon
by black holes is mainly due to the existence of null circular geodesics \cite{c6}.
On the other side of theoretical studies,
the photonsphere is useful in determining the effective
length of the hair above the hairy black hole horizon \cite{s1,s2,s3,s4,s5,s6}.
And it was also shown that the photonsphere provide the fast way
to circle a black hole \cite{s7,s8,s9}.
In addition, it was found that stable photonspheres of compact star can trigger nonlinear instabilities
to massless field perturbations \cite{us1,us2,us3,us4,us5,us6,us7}.
And the characteristic resonances of black holes are
related to unstable circular null geodesics \cite{ad1,r1,r2,r3,r4,r5,r6,r7,r8}.

The horizonless star with photonspheres is usually
called regular ultra-compact star.
Bounds on the compactness and photonsphere radii were studied.
In the case of positive energy-momentum trace, the lower bound on the
compactness parameter of horizonless ultra-compact star
was studied in \cite{b1}. And the discussion was also extended to
the regular ultra-compact star with negative energy-momentum trace \cite{b2}.
In the black hole background, it was shown that the photonsphere radius
has an upper bound expressed in terms of the
total ADM mass of the spacetime \cite{b3}.
Along this line, we try to examine whether there are similar bounds on
photonspheres of horizonless ultra-compact stars.

This paper is planed as follows.
In section II, we introduce the regular compact star
with photonspheres in the asymptotically flat gravity.
In section III, we analytically obtain upper bounds on
the radius of regular compact star photonsphere.
And the last section is devoted to our main results.

\section{The gravity model of regular compact stars}

We consider static spherically symmetric horizonless ultra-compact star
which possesses null circular geodesics. In Schwarzschild coordinates,
the compact star geometry are described by the line element \cite{c2,s1,s3}
\begin{eqnarray}\label{AdSBH}
ds^{2}&=&-g(r)e^{-2\chi(r)}dt^{2}+g^{-1}dr^{2}+r^{2}(d\varphi^2+sin^{2}\varphi d\phi^{2}).
\end{eqnarray}
The solutions $\chi(r)$ and $g(r)$
are functions of the radial coordinate r.
Regularity of the gravity at the center requires \cite{b1,b2}
\begin{eqnarray}\label{AdSBH}
g(r\rightarrow 0)=1+O(r^2)~~~~~~and~~~~~~\chi(0)<\infty.
\end{eqnarray}

The spacetime at the infinity is asymptotically flat,
which is characterized by
\begin{eqnarray}\label{AdSBH}
g(r\rightarrow \infty)=1~~~~~~and~~~~~~\chi(r\rightarrow \infty)=0.
\end{eqnarray}

We state that these spherically symmetric stars could be solutions of
a perfect fluid coupled to the gravity background.
According to Einstein equations $G^{\mu}_{\nu}=8\pi T^{\mu}_{\nu}$,
the anisotropic energy momentum tensor is $T^{\mu}_{\nu}
=diag\{\frac{1-g-rg'}{8\pi r^2},~\frac{-1+g+rg'-2rg\chi'}{8\pi r^2},
~-\frac{2g(\chi'-r\chi'^2+r\chi'')+g'(3r\chi'-2)-rg''}{16\pi r},
~-\frac{2g(\chi'-r\chi'^2+r\chi'')+g'(3r\chi'-2)-rg''}{16\pi r}\}$.
Here we define $T^{t}_{t}=-\rho$, $T^{r}_{r}=p$ and $T^{\varphi}_{\varphi}=T^{\phi}_{\phi}=p_{T}$
as the energy density, the radial pressure and the tangential pressure respectively \cite{s3,t1}.
And the equations of metric solutions
can be expressed as
\begin{eqnarray}\label{BHg}
g'=-8\pi r \rho+\frac{1-g}{r},
\end{eqnarray}
\begin{eqnarray}\label{BHg}
\chi'=\frac{-4\pi r (\rho+p)}{g},
\end{eqnarray}

The gravitational mass $m(r)$ within a sphere of radius r is given by the integration
\begin{eqnarray}\label{AdSBH}
m(r)=\int_{0}^{r}4\pi r'^{2}\rho(r')dr'.
\end{eqnarray}
And the metric solution can be put in the form \cite{b2}
\begin{eqnarray}\label{BHg}
g=1-\frac{2m(r)}{r}.
\end{eqnarray}

According to (6), a finite mass configuration is characterized by \cite{b1}
\begin{eqnarray}\label{AdSBH}
r^{3}\rho(r)\rightarrow 0~~~~~~as~~~~~~r\rightarrow\infty.
\end{eqnarray}

\section{Upper bounds on radii of compact star photonspheres}

In this part, we prove a generic upper bound on the photonsphere of compact stars.
We firstly follow the analysis in \cite{c2,s3,us2} to obtain the characteristic equation
of the photonsphere in the spherically symmetric compact star background.
The conservation equation $T^{\mu}_{\nu;\mu}$ has only one nontrivial component
\begin{eqnarray}\label{BHg}
T^{\mu}_{r;\mu}=0.
\end{eqnarray}

Substituting equations (4) and (5) into (9), we arrive at
\begin{eqnarray}\label{BHg}
p'(r)=\frac{1}{2rg}[(3g-1-8\pi r^2p)(\rho+p)+2gT-8gp]
\end{eqnarray}
with $T=-\rho+p+2p_{T}$ as the trace of the energy momentum tensor.

With the pressure function $P(r)=r^2p$,
the relation (10) can be transformed into
\begin{eqnarray}\label{BHg}
P'(r)=\frac{r}{2g}[N(\rho+p)+2gT-4gp],
\end{eqnarray}
where $N=3g-1-8\pi r^2p$.

We assume that the matter fields satisfy the dominant energy condition
\begin{eqnarray}\label{BHg}
\rho\geqslant |p|,~|p_{T}|\geqslant 0.
\end{eqnarray}

According to (8) and (12), the pressure function $P(r)$ has the asymptotical behavior
\begin{eqnarray}\label{BHg}
P(r\rightarrow 0)\rightarrow 0~~~~~~and~~~~~~rP(r\rightarrow \infty)\rightarrow 0.
\end{eqnarray}

With relations (2), (3) and (13), the radial
function $N(r)$ satisfies
\begin{eqnarray}\label{BHg}
N(r=0)=2~~~~~~and~~~~~~N(r\rightarrow\infty)\rightarrow2.
\end{eqnarray}

In the spherically symmetric spacetime, the photonsphere is characterized by
\begin{eqnarray}\label{BHg}
V_{r}=E^{2}~~~~~~and~~~~~~V_{r}'=0,
\end{eqnarray}
where $V_{r}$ is the effective radial potential that governs the null trajectories in the form
\begin{eqnarray}\label{BHg}
V_{r}=(1-e^{2\chi})E^{2}+g\frac{L^2}{r^2}.
\end{eqnarray}
Here E is the conserved energy and L is the conserved angular momentum
in accordance with the independence
of the metric (1) on both t and $\phi$.

Substituting Einstein equations (4) and (5) into
(15) and (16), the photonsphere is determined
by the characteristic relation \cite{b1}
\begin{eqnarray}\label{BHg}
N(r_{\gamma})=3g(r_{\gamma})-1-8\pi (r_{\gamma})^2p=0.
\end{eqnarray}
The roots of (17) correspond to the discrete radii of the null circular geodesics.
For the case of a Schwarzschild black hole, there
is $g=1-\frac{2M}{r}$, $\chi=0$ and $p=\frac{-1+g+rg'-2rg\chi'}{8\pi r^2}=0$,
which yields the familiar $r_\gamma=3M$.

We define $r_{\gamma}^{out}$ as the outermost photonsphere of the regular ultra-compact objects.
From Eqs. (14) and (17), one deduces that the outermost photonsphere of the spherically symmetric horizonless
ultra-compact objects satisfies the relation \cite{b1,b2,us5}
\begin{eqnarray}\label{BHg}
N'(r=r_{\gamma}^{out})\geqslant0.
\end{eqnarray}
We point out that spatially regular horizonless spacetimes usually possess an even number of photonspheres
and the degenerate case of $N'(r=r_{\gamma})=0$ may be characterized by odd number of photonspheres \cite{us4,us5}.

From relations (4), (11) and (17), we get the function \cite{b1,b2,us5}
\begin{eqnarray}\label{BHg}
N'(r=r_{\gamma})=\frac{2}{r_{\gamma}}[1-8\pi r_{\gamma}^2(\rho+p)].
\end{eqnarray}
Putting (19) into (18), we obtain the inequality
\begin{eqnarray}\label{BHg}
8\pi (r_{\gamma}^{out})^2(\rho+p)\leqslant 1
\end{eqnarray}
at the outermost photonsphere of the ultra-compact star.

With (12), (17) and (20), we obtains the relations \cite{b2,b3}
\begin{eqnarray}\label{BHg}
g(r_{\gamma}^{out})=\frac{1+8\pi(r_{\gamma}^{out})^2p}{3}=\frac{1}{3}+\frac{8\pi(r_{\gamma}^{out})^2p}{3}\leqslant
\frac{1}{3}+\frac{4\pi(r_{\gamma}^{out})^2(\rho+p)}{3}\leqslant\frac{1}{2}+\frac{1}{6}=\frac{1}{2}.
\end{eqnarray}

Considering the relations (7) and (21), we have
\begin{eqnarray}\label{BHg}
\frac{2m(r_{\gamma}^{out})}{r_{\gamma}^{out}}\geqslant \frac{1}{2}.
\end{eqnarray}

So we obtain an upper bound on the radii of photonspheres
\begin{eqnarray}\label{BHg}
r_{\gamma}\leqslant r_{\gamma}^{out}\leqslant 4m(r_{\gamma}^{out})=4\int_{0}^{r_{\gamma}^{out}}4\pi r'^{2}\rho(r')dr'\leqslant 4\int_{0}^{\infty}4\pi r'^{2}\rho(r')dr'=4M.
\end{eqnarray}

We also consider the matter field configurations with the negative isotropic trace
\begin{eqnarray}\label{BHg}
T=-\rho+3p<0~~~~~~or~~~~~~\rho>3p.
\end{eqnarray}

According to (17), (20) and (24), we obtain the series of relations \cite{b2}
\begin{eqnarray}\label{BHg}
g(r_{\gamma}^{out})=\frac{1+8\pi(r_{\gamma}^{out})^2p}{3}\leqslant \frac{1+2\pi(r_{\gamma}^{out})^2(\rho+p)}{3}\leqslant\frac{1+\frac{1}{4}}{4}=\frac{5}{12}.
\end{eqnarray}

Taking cognizance of relations (7) and (25), we arrive at the inequality
\begin{eqnarray}\label{BHg}
\frac{2m(r_{\gamma}^{out})}{r_{\gamma}^{out}}\geqslant \frac{7}{12}.
\end{eqnarray}

So a stronger upper bound can be obtained for $T<0$ as
\begin{eqnarray}\label{BHg}
r_{\gamma}\leqslant r_{\gamma}^{out}\leqslant \frac{24}{7}m(r_{\gamma}^{out})\leqslant \frac{24M}{7}.
\end{eqnarray}

We mention that the spherically symmetric asymptotically flat black hole photonsphere
has an upper bound $r_{\gamma}\leqslant 3M$ \cite{b3}. In the black hole, there is a condition $N(r_{H})\leqslant0$
at the horizon $r_{H}$, which play an important role in the analysis.
In this horizonless compact star, we have no such relation and instead there is
$N(0)=2>0$ at the center.For this reason, we cannot simply follow the analysis of black hole photonsphere
in \cite{b3} to obtain the bound on the regular star photonsphere.
We believe it is interesting to further search for stronger
upper bounds on the compact star photonsphere and
examine whether there is regular star
photonsphere, which can saturates the bound (23) and (27) \cite{j1,j2,j3,j4,j5,j6,j7,j8,j9}.
It is known that one way to construct hairy compact objects is
enclosing the compact objects in a box \cite{j10,j11,j12,j13,j14,j15,j16,j17}.
So it is also very interesting to examine the photonsphere radius bound
in the confined gravity.

\section{Conclusions}

We studied photonspheres in the background of
horizonless asymptotically flat ultra-compact stars.
We showed that the radius of the compact star photonsphere is
bounded from above by $r_{\gamma}\leqslant 4M$,
where $r_{\gamma}$ is the radius of the photonsphere
and M is the total ADM mass of the spacetime.
In the case of negative isotropic trace,
we obtained a stronger upper bound in the form
$r_{\gamma}\leqslant \frac{24M}{7}$.
The analysis in this work can be applied to the
general gravity model considering
the matter fields' backreaction on the compact star
in various asymptotically flat gravity theories.

\begin{acknowledgments}

We would like to thank the anonymous referee for the
constructive suggestions to improve the manuscript.
This work was supported by the Shandong Provincial 
Natural Science Foundation of China under Grant
No. ZR2018QA008.

\end{acknowledgments}

\end{document}